\documentclass{aa}  

\usepackage{graphicx}	
\usepackage{amsmath}	
\usepackage{amssymb}	
\usepackage{newtxtext,newtxmath}
\usepackage{multirow}
\usepackage{lscape}
\usepackage{makecell}
\usepackage{subfigure}
\usepackage{float}
\usepackage{newtxtext,newtxmath}
\usepackage{footmisc}

\usepackage{txfonts}
\begin{document}

   \title{Testing X-ray Periodicity and Long-Term Trend in PG 1553+113 via Targeted \textit{Swift}-XRT Monitoring}

   \author{P. Pe\~nil\inst{1}
          \and
          N. Torres$-$Alb\`a\inst{2, 3} 
          \and
          L. Marcotulli\inst{1,4}
          \and
          A. Dom\'inguez\inst{5}
          \and
          M. Ajello\inst{1}
          \and
          A. Rico\inst{1}
          \and
          S. Buson\inst{4,6}
          \and 
          S. Adhikari\inst{1}
          }

   \institute{
            Department of Physics and Astronomy, Clemson University, Kinard Lab of Physics, Clemson, SC 29634-0978, USA\\
            \email{ppenil@clemson.edu}
        \and
            Department of Astronomy, University of Virginia, P.O. Box 400325, Charlottesville, VA 22904, USA \\
            \email{nuria@virginia.edu}
        \and
             GECO Fellow\\
        \and
            IPARCOS and Department of EMFTEL, Universidad Complutense de Madrid, E-28040 Madrid, Spain\\
        \and
            Deutsches Elektronen-Synchrotron DESY, Platanenallee 6, 15738 Zeuthen, Germany\\
            \email{lea.marcotulli@desy.de}        
        \and 
             Julius-Maximilians-Universit\"at W\"urzburg, Fakultät f\"ur Physik und Astronomie, Emil-Fischer-Str. 31, D-97074 W\"urzburg, Germany\\      
        }

   \date{Received XX, XX; accepted April 6, 2026}

 
  \abstract
   {PG~1553+113 is the blazar with the most-significantly detected periodic patter in its multiwavelength (MWL) emission, making it one of the most promising candidates for hosting a supermassive black hole binary.}
   {However, the presence of this periodic behavior in the X-ray band remains under debate, largely due to the lack of continuous monitoring. This has led to differing conclusions in previous studies. In addition, we aim to examine whether the recently identified linear long-term trends in the $\gamma$-ray and optical bands also exist in the X-ray regime.}
   {Here, we evaluate the $\sim$2.1-year period in the X-ray light curve of PG~1553+113 using two dedicated monitoring campaigns with \textit{Swift}-XRT and UVOT, guided by predictions of future oscillation phases. We also examine whether the long-term trend is present in X-rays, the potential periodic behavior of the X-ray power-law photon index, and its potential correlation to the X-ray flux.}
   {As a result, we find tentative evidence for a correlation between the predicted high-emission states in the $\gamma$-ray band and those observed in the X-ray and UV bands. Therefore, we do not find a strong evidence of the same periodic pattern in X-ray. In addition, we find that the X-ray light curve is consistent with the presence of a long-term linear trend, in agreement with those previously reported in $\gamma$-ray, optical, and radio.}
   {Overall, these results indicate that the X-ray emission is likely to share the same long-term behavior observed in the $\gamma$-ray and optical bands. Nevertheless, the pronounced stochastic variability that characterizes the X-ray light curve limits our ability to draw firm conclusions regarding the presence of the periodic behavior.}


   \maketitle
%

\section{Introduction}
PG 1553+113 is a BL Lacertae object (hereafter BL Lac) and one of the most extensively studied blazars due to its distinctive long-term variability across multiple wavelengths \citep[][]{2007ApJ...654L.119A}. Located at a redshift of $z\approx$0.433 \citep[][]{redshift_pg1553_jones_johson, redshift_pg1553_jones}, its emission spans the entire electromagnetic spectrum, from radio to very high-energy $\gamma$ rays \citep[e.g.][]{ackermann_pg1553, penil_mwl_pg1553, magic_pg1553_2024}. It is classified as a high-frequency–peaked BL Lac object (HBL), meaning that the peak of its synchrotron spectral energy distribution lies in the UV–X-ray band \citep[e.g.][]{blac_xray_1995, sed_classification_2010}. As a consequence, the X-ray emission of PG 1553+113 is dominated by synchrotron radiation from the highest-energy electrons in the jet.

A particularly intriguing feature of PG 1553+113 is the periodic behavior observed in its $\gamma$-ray emission. Early analyses of \textit{Fermi}-LAT data reported a significant periodicity of approximately 2.1 years, with the number of identified cycles ranging from approximately four to seven, depending on the time span of the analyzed light curve \citep[e.g][]{ackermann_pg1553, penil_2020, alba_ssa}. Interestingly, this periodic signature is not limited to the $\gamma$-ray band; similar periodicities have also been detected in the optical, UV, and radio bands, with approximately four (radio) to eight (optical) cycles identified depending on the data available for each energy band \citep[e.g.][]{ackermann_pg1553, penil_mwl_pg1553, magic_pg1553_2024}, pointing toward a common physical mechanism modulating the emission across these wavelengths.

In addition to this periodicity, a new long-term trend has recently been identified. Multi-wavelength observations have revealed a gradual, approximately linear increase in flux over time in the $\gamma$-ray, UV, optical, and radio bands \citep{penil_mwl_pg1553}. The slope of this increasing trend appears consistent across all these energy bands, suggesting a coherent long-term process influencing the entire spectral energy distribution. This trend was proposed by \citet{gao_pg1553_2023} as part of another periodic oscillation of 42 years. To explore the potential origin of this trend, \citet{sagar_pg1553} analyzed more than a century of historical optical data, uncovering a possible secondary periodicity of $\sim$22 years. This finding implies that the variability of PG 1553+113 may be governed by more than one characteristic timescale.

While the $\sim$2.1-year periodicity is well established in the $\gamma$-ray, UV, and optical bands, the behavior in the X-ray regime remains more ambiguous. Several studies have proposed different periodicities in the \textit{Swift}-XRT X-ray light curve of PG 1553+113. For instance, \citet{huang_pg1553_xperiod} reported a 2.1-year period, consistent with that observed in the higher and lower energy bands. In contrast, \citet{penil_mwl_pg1553} found evidence for a shorter 1.5-year period with a significance of approximately 2.8$\sigma$\footnote{\label{fn:global}The global significance is $\sim$0$\sigma$, using the methodology presented in \citet{penil_mwl_pg1553,penil_24candidates_2025}}, as well as a secondary period of 2.3 years, also at 2.8$\sigma$\footref{fn:global} significance. In \citet{aniello_pg1553_xray_2024}, a period of 1.4 years was observed. These discrepancies suggest that the \textit{Swift}-XRT X-ray emission is likely governed by additional or distinct variability mechanisms compared to other wavebands. In particular, the X-ray light curve may be dominated by strong stochastic variability, which can either obscure an underlying periodic pattern or render the emission effectively consistent with red-noise-dominated behavior, thereby reducing the detectability of a coherent periodic signal \citep[e.g.,][]{vaughan_red_noise, marscher_regions_jet_2014, haocheng_harder_brighter_2015, petropoulou_radiative_cooling_2018}.

Understanding these differences is essential for advancing our knowledge of the variability patterns of PG 1553+113. In particular, unraveling the connection between the X-ray and $\gamma$-ray variability is key to constraining models of emission processes, jet structure, and central engine dynamics \citep[e.g.,][]{2013ApJ...774...18Z, 2025MNRAS.536.3242O}. 

This paper presents a comprehensive analysis of the long-term X-ray variability of PG 1553+113 and explores its correlation with the observed $\gamma$-ray and UV activity. Our goal is to offer new insights into the temporal properties of this blazar, contributing to a better understanding of its broadband variability.

To achieve this, we conducted a targeted observational campaign using \textit{Swift}-XRT and UVOT. Specifically, we designed two monitoring programs aimed at capturing the X-ray behavior during expected high-emission phases, as predicted by extrapolating the period of $\sim$2.1-year modulation previously observed in MWL data. These observations were strategically scheduled to coincide with predicted maxima in the $\gamma$-ray light curve, allowing us to investigate whether a similar periodic pattern is present in the X-ray band and to assess the degree of cross-band coherence. Additionally, these observations provide an opportunity to better characterize any long-term trends in the X-ray emission, analogous to those reported in the MWL variability. 

The structure of this paper is organized as follows. In Section \ref{sec:theoretical_models}, we introduce the different theoretical models proposed to explain the periodic emission of PG 1553+113. Section \ref{sec:campaings} presents the \textit{Swift} observational campaigns aimed at evaluating the presence of the 2.1-year modulation in the X-ray band. In Section \ref{sec:data_analysis}, we describe the data reduction and analysis procedures used to construct the MWL light curves. Section \ref{sec:methodology} outlines the different techniques and methods employed to investigate the long-term variability of the light curves. The results obtained from the application of these methodologies are presented in Section \ref{sec:results}. A discussion of these results is provided in Section \ref{sec:discussion}. Finally, the main conclusions of this work are summarized in Section \ref{sec:summary}.  

\section{Theoretical Models}\label{sec:theoretical_models}
Several models have been proposed to explain the MWL periodic variability of PG 1553+113. Most of these models are based on the supermassive black hole binary (SMBHB). A commonly interpretation for the observed $\sim$2.1-yr period is a SMBHB at sub-parsec separations, whose orbital dynamics may modulate the emission through accretion-flow perturbations or geometric changes in the jet orientation \citep[e.g.][]{ackermann_pg1553, caproni_pg1553_2017, sobacchi_binary_2017, tavani_pdm_pg_1553, stefan_pg1553_2024}. Among these possibilities, jet precession is one of the most widely discussed mechanisms \citep[][]{camenzind_jet}. In this model, the secondary black hole exerts a torque on the accretion disk, inducing precession of the disk and the relativistic jet. The resulting periodic changes in the viewing angle can then produce strong flux variations through Doppler boosting \citep[][]{villata_helical_jet}.

In the context of PG 1553+113, \citet{gao_pg1553_2023} proposed a model based on the precession of the jet (without specifying the potential cause of such precession), where the jet rotates with constant angular velocity around an axis, causing the Doppler factor of the jet to vary over time, leading to periodic changes in flux. This scenario reproduces both the observed periodicity and the rising trend observed in the $\gamma$-ray light curves. \citet{caproni_pg1553_2017} proposed two different models. The first model is based on a single jet model; the jet is launched by the primary black hole, and its precession is driven by the gravitational influence of the secondary black hole. In the alternative model, both black holes launch their own jets, and the observed variability results from the interaction and combined emission of these precessing jets. Similarly, \citet{sobacchi_binary_2017} also uses the periodicity produced by the jet's precession due to the orbital motion of the jet-emitting SMBH around its companion. 

\citet{tavani_pdm_pg_1553} explains the periodic oscillation based on an orbiting binary system. In addition to the main peaks of the periodic oscillations, the study identifies secondary peaks, named ``twin peaks'' that occur symmetrically around the main peaks. Two scenarios are proposed to interpret these observations. In the first one, we are observing the jet of the primary SMBH, and the secondary black hole induces additional instabilities in the primary's jet during its orbital motion, leading to the observed twin peaks. In the second model, the secondary black hole launches its own precessing jet, contributing to the twin peaks observed in the emission profile.  

Precession can also arise from other mechanisms. For instance, in single SMBH systems, Lense-Thirring precession induced by misalignment between the black hole spin axis and the disk angular momentum vector may lead to periodic modulation of the jet orientation \citep[][]{franchini_lense, zanazzi_lense}. 

In binary systems, tidal forces from the secondary black hole can further enhance accretion rate oscillations by periodically disturbing the outer disk. These modulations may imprint themselves on the jet's emission \citep{gracias_modulation_disk}. In this context, \citet{stefan_pg1553_2024} suggested that the companion object triggers a Keplerian orbital modulation in the accretion rate onto the primary SMBH. This interaction leads to periodic hydrodynamic variability, characterized by repeated enhancements in the primary’s accretion disk activity and the corresponding emission from its relativistic jet. 

The observed long-term trend and the secondary period of $\sim$22 years were explained in \citet{sagar_pg1553}. In that work, the dual periodicities reported for PG 1553+113 were interpreted within a binary SMBH framework: the shorter $\sim$2.1-year period was attributed to the orbital motion of the SMBH pair, while the longer $\sim$22-year modulation was associated with the dynamics of an over-dense structure (named “lump”) in the circumbinary accretion disk \citep[][]{orazio_lump_definition}. These lumps are predicted by hydrodynamic simulations of circumbinary disks \citep[][]{WS2022+}, which show that tidal interactions with the binary can drive the formation of non-axisymmetric density enhancements. Additional studies, such as \citet{farris_2014}, showed that such disks can develop long-lived, rotating overdensities that exert variable torques on the binary system and modulate the inflow of matter through the disk cavity. The interaction between the binary and the lump can result in periodic enhancements in accretion rate onto the binary system, which in turn can produce large-scale modulations in the emitted flux.

These previous scenarios imply different expectations for the MWL variability. In geometric models, such as jet precession, the modulation is primarily produced by periodic changes in the viewing angle and, therefore, in the Doppler boosting factor \citep[][]{camenzind_jet, villata_helical_jet, rieger_2004}. In this case, bands produced in closely related jet regions are expected to show coherent long-term variability, with small lags, although the modulation amplitude may remain energy dependent \citep[][]{villata_helical_jet,rieger_2004}. Within this framework, the optical/UV and $\gamma$-ray emission may display similar recurrence patterns \citep[][]{magic_pg1553_2024, penil_mwl_pg1553}, whereas the radio emission can respond later because it is often produced farther downstream in a more extended and partially self-absorbed region of the flow \citep{max-moerbeck2014}. 

By contrast, in accretion-driven binary scenarios the orbital motion modulates the mass supply, thus, the power injected into the jet through periodic perturbations of the accretion flow \citep[][]{lai_2022, stefan_pg1553_2024}. In such models, the resulting MWL signal may show broader flares, band-dependent amplitudes, and possible delays associated with the propagation of disturbances from the disc to the jet-emitting regions \citep[][]{lai_2022}. In addition, circumbinary-disc models based on the ``lumps'' can results in lags between the different bands \citep[][]{farris_2014,lai_2022}. 

For PG~1553+113, these expectations are especially relevant because the available MWL data indicate that the optical variability is strongly correlated with the $\gamma$-ray modulation \citep[e.g.,][]{ackermann_pg1553, penil_mwl_pg1553}, and is consistent with no lag, whereas the radio emission appears correlated but delayed, and the X-ray behavior is less clearly correlated with the long-term $\gamma$-ray oscillation \citep{ackermann_pg1553, penil_mwl_pg1553}. Overall, this phenomenology is broadly consistent with a scenario in which at least part of the modulation is geometric and jet-related \citep[][]{madero_dominguez_2026}, while additional accretion-driven effects may also contribute to shaping the long-term variability \citep{farris_2014, sagar_pg1553}.

\section{\textit{Swift} Observations}\label{sec:campaings}
The periodic nature of the MWL emission can be considered robust, as it has been confirmed by several independent studies \citep[][]{ackermann_pg1553, magic_pg1553_2024, penil_mwl_pg1553, stefan_pg1553_2024}. However, as noted previously, the periodicity in the X-ray band remains a subject of debate due to the conflicting results reported across different analyses \citep[][]{huang_pg1553_xperiod, penil_mwl_pg1553, aniello_pg1553_xray_2024}.

In the literature, periodic behavior in blazar emissions is typically classified into two categories: quasi-periodic oscillations (QPOs) and strict periodicity. These classifications are not only observationally motivated but also mathematically distinct. QPOs refer to variability patterns that show a preferred timescale of oscillation but with a significant stochastic (unpredictable) element. As a result, while a characteristic timescale may be identified, precise prediction of future oscillations is inherently uncertain. On the other hand, strict periodicity implies a deterministic pattern, allowing for the reliable prediction of future behavior. Given the inconsistency among X-ray studies, the variability observed in this band for PG 1553+113 is more appropriately characterized as a QPO, rather than a strictly periodic signal, while at all other wavebands a strict periodicity has been confirmed \citep[][]{ackermann_pg1553, penil_mwl_pg1553}.

To address this ambiguity and test whether the X-ray emission follows a genuinely periodic pattern, we developed two dedicated observing campaigns with the \textit{Swift}-XRT instrument. We designed these campaigns to specifically target epochs when a high X-ray flux would be expected if the emission follows the same periodicity observed in the $\gamma$-ray band (with a period of $\sim$2.1 years). By focusing on predicted high states based on this periodic model, our approach allows for a direct test of phase coherence and recurrence in the X-ray regime. Successful detection of X-ray peaks aligned with the predicted epochs would strengthen the case for a common physical origin driving the periodicity across energy bands, while deviations would support the interpretation of the X-ray variability as stochastic or independently modulated.

\subsection{Observational Campaigns and Archival Data}\label{sec:obser_camp}

The first observational campaign (Program ID: 1821168, PI: Peñil, 12 observations) took place during Cycle 18 of \textit{Swift}, coincident with the expected peak of emission in March 2023 \citep[$\pm$1 month, based on our analysis of the \textit{Fermi} data][]{alba_ssa}. We monitored PG 1553+113 for a total of four months around its predicted high state. We proposed to monitor the source bi-weekly for one month, then once a week for two months (i.e., a month on each side of the expected peak position), and then go back to monitoring bi-weekly for another month. This results in a total of 12 observations, with a duration of 1\,ks. 

The second campaign (Program ID: 2124062, PI: Peñil, 24 observations) took place during \textit{Swift} Cycle 21, coincident with an expected peak of emission in May 2025 \citep[$\pm$1 month, based on our analysis of the \textit{Fermi} data][]{alba_ssa}. In this case, we monitored PG 1553+113 weekly for three months (April, May, and June), totaling 12 observations, aligning with the emission peak and its uncertainty. During other observable months (July-September 2025, January 2025-February 2026), \textit{Swift} monitored the source bi-weekly. 

Additionally to our observational campaigns, PG 1553+113 has been observed by \textit{Swift} a large number of times, with a total of 555\footnote{As of November 10th 2025.} observations taken between October 18th 2014 and September 25th 2025. The majority of these observations contain event files for both photon counting (PC) and window-timing (WT) modes; with a total of 404 and 420 event files respectively. We include all available observations in our initial analysis. 

\section{Data Analysis}\label{sec:data_analysis}

Both the \textit{Swift}-XRT and \textit{Swift}-UVOT data were analyzed using the \textit{Swift} Analysis Pipeline for Lightcurve Extraction (SAPLE)\footnote{SAPLE GitHub link: \url{https://github.com/leamarcotulli/saple}}. \textit{Fermi}-LAT $\gamma$-ray data were retrieved from the public light-curve repository \citep[][]{fermi_repository}\footnote{\url{https://fermi.gsfc.nasa.gov/ssc/data/access/lat/LightCurveRepository/about.html}}. The detailed analysis procedures for each instrument are described in the following subsections.

\subsection{XRT Data Analysis}\label{sec:xray_flux}

We use data from \textit{Swift}-XRT \citep[X-ray Telescope,][]{Burrows2005}, and extract spectra in the 0.3$-$10~keV band, as is the default in SAPLE. The core of the pipeline is composed of a total of four codes, which we run as indicated in the documentation. 

The first code, \texttt{xrtpipeline\_run.py} runs the official HEASoft\footnote{HEASoft version V6.35. The Swift XRT CALDB version is 20240522} command \texttt{xrtpipeline} on all observations (both on the photon counting, PC, and the windowed time, WT, observations). As a second step, we open one of the XRT images of the target and we select a default circular source and background regions, of 50'' and 90'' respectively, for PC events, and 50'' and 50'' for WT events. The source region is centered at the target location, while the background region is chosen in a source free region close by the target. 

The second code, \texttt{xrt\_make\_image.py}, produces images of all observations, together with the corresponding source and background regions, used for spectral extraction. To ensure the source is always at the center of the source circle, the code automatically centroids the circle and saves a new source region for each observation. These images are then visually inspected for evident issues in each observation (e.g., the source region is missing the source, the background region contains contaminant sources, empty event files, etc.). A total of three observations are discarded at this point due to the presence of artifacts (see Appendix \ref{appendix:discards} for details, as well as examples).

After removing problematic observations, the third code, \texttt{xselect\_run.py}, is run.This code creates spectra for both the source and background regions selected. Finally, the fourth code, \texttt{xspec\_pl\_fit.py}, performs the following steps for each observation: 1) using the \texttt{xrtmkarf} task, it produces an ancillary response file and, using \texttt{grppha}, it assigns the corresponding background, ancillary response file (ARF) and response matrix file (RMF\footnote{\texttt{swxpc0to12s6\_20210101v016.rmf} for PC and \texttt{swxwt0to2s6\_20210101v017.rmf} for WT.}) to the source spectra; 2) it rebins the spectra using the optimal binning scheme provided by \texttt{ftgrouppha}; 3) it fits a redshifted power-law model corrected by Galactic absorption \citep[\texttt{tbabs*zpowerlaw} in XSPEC, using the \texttt{wilm} abundances,][]{Wilms2000} to the obtained spectrum using Cash statistics (C-stat), and extracts the following information: count-rate, observed and intrinsic flux \citep[corrected for galactic absorption, taken to be $3.61\times10^{20}$~cm$^{-2}$,][]{HI4PI2016} and associated uncertainties, photon index and associated uncertainties, and a goodness of fit estimator (i.e. C-stat over degrees of freedom).

Out of the 817 analysed archival XRT observations, 34 have non-positive counts, 154 do not have a converge fit, and 41 have error estimations that do not converge. All of these issues are tracked back to a lack of signal-to-noise ratio\footnote{Most observations have both a PC and WT, often with very different exposure times, which can be shorter than 10s.} in the WT observations. We end up with a total of 585 usable data points after the analysis. The XRT light curve is presented in the second panel of Fig. \ref{fig:mwl_pg1553}, highlighting the two predicted emission peaks targeted with the monitoring campaigns. The photon index data is used in Sect. \ref{sec:results}.

\subsection{UVOT Data Analysis}

We use data from \textit{Swift}-UVOT (Ultraviolet and Optical Telescope)\footnote{\url{https://www.swift.ac.uk/about/instruments.php}}, considering all available filters\footnote{\url{https://www.swift.ac.uk/analysis/uvot/filters.php}} for each observation. We use the SAPLE codes dedicated to UVOT analysis as follows. To start, we run \texttt{uvot\_src\_bkg\_regions.py} to create source and background regions (5'' and 30'' circles, respectively) for all observations, as well as create images of the FOV. In a process similar to that of the XRT pipeline, we visually inspect all images to discard problematic observations. Specifically, UVOT images can be affected by imperfect tracking, causing streaks in the images/event files, resulting in incorrect photometry. A total of 28 observations are affected, and removed from the analysis (see Appendix \ref{appendix:discards} for details and examples). 

After that, we run \texttt{uvotsource\_run.py}, which excecutes the standard \texttt{uvotsource} task to extract the source and background magnitude for each observation in all available filters. Finally, we run \texttt{uvotsource\_extract\_flux.py} which performs the following tasks: 1) corrects the AB magnitudes for Galactic extinction following the prescription in \cite{Romning2009}. The extinction values (E(B-V)) for the source are taken from \cite{Willingale2013}\footnote{\url{https://www.swift.ac.uk/analysis/nhtot/index.php}}; 2) transforms the corrected magnitudes and corresponding uncertainties to flux values. We choose to perform the analysis in magnitudes.

We start from 3052 photometric data points (considering all filters available per observation). Removing 28 observations results in the loss of 152 photometric points, which leaves us with a total of 2900. As an example, data for the UV band (from the filter "uvv") is shown in the bottom panel of Fig. \ref{fig:mwl_pg1553}.

\subsection{\textit{Fermi}-LAT Data}
\begin{figure}
        \centering
        \includegraphics[scale=0.15]{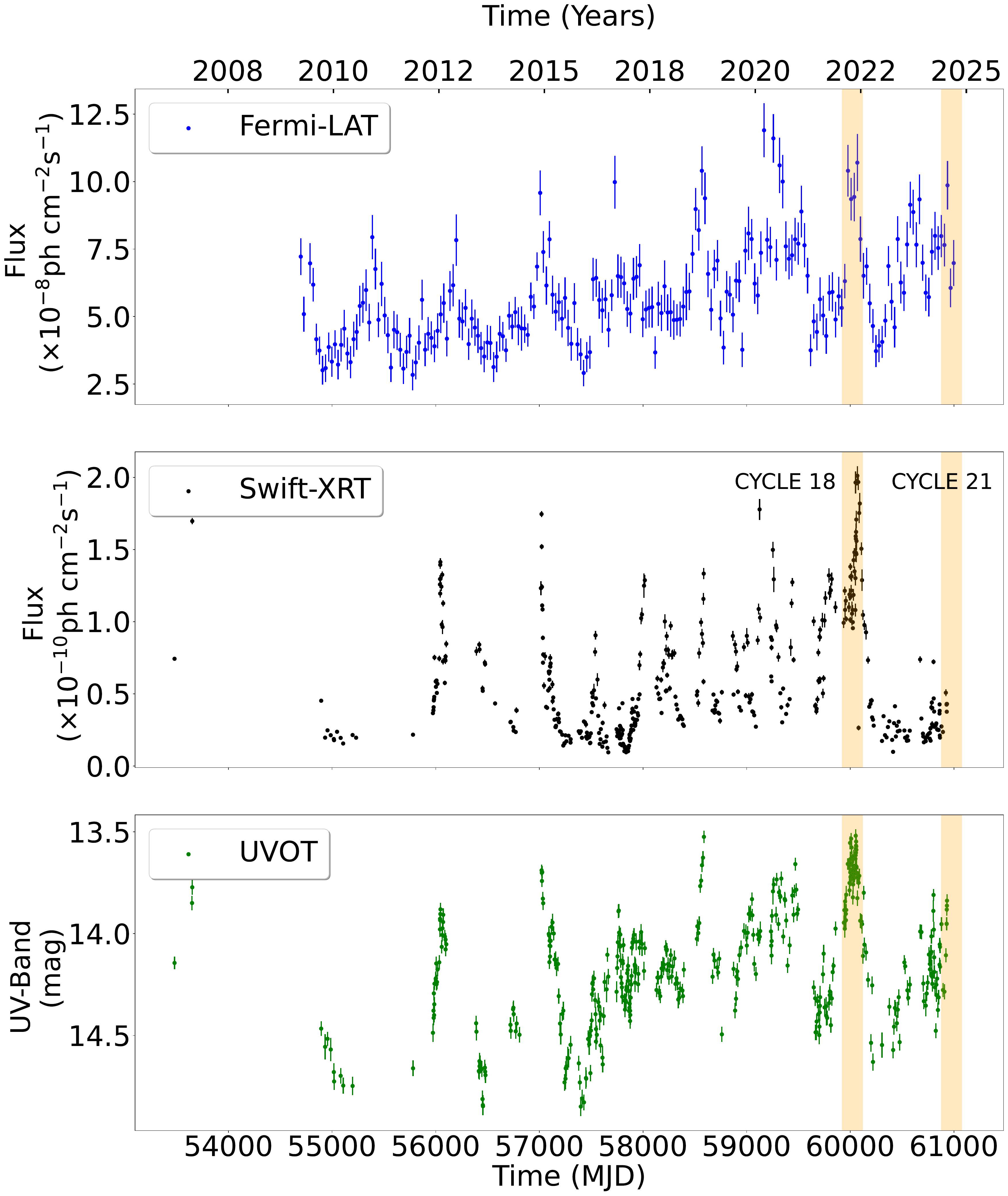}
        \caption{Multiwavelength light curves of PG 1553+113 employed in this study. The orange vertical bands indicate the approximate location of the \textit{Swift} monitoring campaigns conducted to track the predicted high-emission states. \textit{Top:} \textit{Fermi}-LAT light curve, 30-day binned. \textit{Center:} \textit{Swift}-XRT light curve, including both PC and WT observations. \textit{Bottom:} UVOT UV band light curve (filter "uvv"). Our analysis is restricted to data from MJD $\sim$55000 onward, in order to avoid potential inconsistencies associated with the large data gap prior to that date \citep[][]{penil_gaps_2025}.} \label{fig:mwl_pg1553}
\end{figure}

We use the $\gamma$-ray data from the open-access \textit{Fermi} LAT Light Curve Repository, which provides comprehensive data covering approximately 18 years of observations by the \textit{Fermi}-LAT. For this study, we have selected the 30-day binned light curves, using a fixed photon index. This binning interval is strategically chosen to facilitate the search for a period of $\approx$2.1 years and the long-term trend. Figure~\ref{fig:mwl_pg1553}, top, presents the \textit{Fermi}-LAT light curve used in this study.

\section{Methodology}\label{sec:methodology}
In this study, we perform three types of variability analyses: periodicity searches, cross-correlation studies, and trend characterization. The specific methodologies applied to each case are described in detail below.

\subsection{Periodicity Search} 
To carry out the periodicity search, we selected methods that are well suited to the specific challenges of the X-ray light curve. In particular, X-ray light curve is characterized by irregular sampling, data gaps, and strong flaring activity, all of which can significantly affect periodicity searches. These effects were systematically studied in \citet{penil_flares, penil_gaps_2025}, where the impact of flares and gaps on the detection and significance of periodic oscillations was quantified. We therefore restricted our analysis to methods whose statistical performance was directly characterized under these observational conditions. Consequently, we did not include other techniques, such as the Weighted Wavelet $Z$-transform \citep[WWZ,][]{wwz}, which was not calibrated in those previous studies.

For these reasons we employed the Generalized Lomb–Scargle Periodogram \citep[GLSP,][]{lomb_gen}, Singular Spectrum Analysis \citep[SSA,][]{ssa_greco, SSA_algorithm}, and Phase Dispersion Minimization \citep[PDM,][]{pdm_stellingwerf}. The GLSP extends the traditional Lomb–Scargle Periodogram \citep[][]{lomb_1976, scargle_1982} by explicitly incorporating measurement uncertainties into the calculation of power. By considering these uncertainties, the GLSP includes the influence of data error on the detection of periodic signals, thereby enhancing the reliability of the period estimates. 

SSA is a method designed to decompose a time series into its fundamental components, enabling the distinction between structured signals and noise. In practice, the method first embeds the light curve into a trajectory matrix constructed from lagged copies of the original series, where the embedding dimension is set by the window length. This matrix is then decomposed through singular value decomposition into a set of components, each representing a different mode of variability. Periodic or quasi-periodic signals are typically associated with a component that shows the oscillatory structure of the light curve, whereas stochastic fluctuations are usually captured by other components. By reconstructing the deterministic component associated with the candidate oscillation, SSA reduces the influence of noise and transient features such as flares, making periodic patterns easier to identify in the original light curve. After isolating the oscillatory component from the stochastic ones, we apply the LSP to the reconstructed signal. This two-step approach improves the robustness of the period search and helps quantify the associated uncertainties \citep{alba_ssa}.

Finally, PDM offers a complementary approach: it tests trial periods by minimizing the scatter (dispersion) of phase-folded data, making it well-suited for detecting non-sinusoidal periodicities that may be missed by Fourier-based techniques, such as LSP. 

The significance of the potential period reported by these methods is assessed by generating 100,000 artificial light curves that replicate the original light curve's characteristics, including its sampling pattern, power spectral density (PSD), and the probability distribution function of the original data. To this end, we apply the method of \citet{emma_lc}. The PSD of each light curve segment is modeled using a power-law function of the form $A*f^{-\beta}$+$C$, where $A$ is the normalization factor, $\beta$ is the spectral index, $f$ denotes frequency, and $C$ represents the Poisson noise level. The model parameters are estimated via Maximum Likelihood Estimation, supported by a Markov Chain Monte Carlo (MCMC) analysis\footnote{Implemented using the \texttt{emcee} Python package}.

\subsection{Correlation}
To analyze correlations in unevenly sampled data, we apply the \textit{z}-transformed Discrete Correlation Function \citep[\textit{z}-DCF,][]{zdfc_alexander}, a method specifically developed to handle irregular time sampling. Unlike the standard Discrete Correlation Function, the \textit{z}-DCF reduces the bias associated with uneven observational cadences. This approach enables us to estimate time lags between the $\gamma$-ray emission and those in the X-ray and UV bands. The significance of the correlations is estimated using the same approach as in the periodicity study. 

\subsection{Trend Characterization}
Finally, to characterize the trend, we use the methodology presented in \citet{penil_2025_trend}\footnote{We use the \texttt{seasonal\_decompose} function of the Python package \texttt{Statsmodels}. The parameters for the function are set as ``Multiplicative'' for the ``Model'' parameter and ``40'' for the ``Period'' parameter.\label{fn:method}}. As was estimated in \citet{penil_mwl_pg1553}, the trend associated with the emission of PG 1553+113 is linear. Consequently, we employ a linear regression\footnote{We use the \texttt{LinearRegression} function of the Python package \texttt{Scikit-learn}, which is optimized specifically for linear fitting}, characterizing the parameters of the linear trend. 

We also use the R-squared (R$^{2}$) criterion as a metric to evaluate how well the trend fits the data. This metric quantifies the proportion of the variance in the dependent variable that can be explained by the independent variable(s) in a regression model. R$^{2}$ values range from 0 to 1, with higher values indicating that the model accounts for a greater portion of the variability observed in the data. However, what constitutes a good R$^{2}$ depends on the context and specific objectives of the analysis. There is no absolute cutoff, as acceptable levels may vary across fields. According to \citet{hair_r2_2011}, R$^{2}$ values of 0.25, 0.50, and 0.75 are typically interpreted as indicating \textit{weak}, \textit{moderate}, and \textit{substantial}, respectively.

\section{Results}\label{sec:results}
As an initial step, we assessed the accuracy of our predictions for the high-emission states of PG 1553+113. As shown in Fig. \ref{fig:mwl_pg1553}, the results are mostly consistent with expectations. During \textit{Swift} Cycle 18, the $\gamma$-ray flux reached its maximum between February and May 2023, coinciding with an X-ray peak observed between April and May 2023. The UVOT data also exhibited a flux maximum during the same period, around April 2023. For \textit{Swift} Cycle 21, the $\gamma$-ray presents an irregular oscillations with,  maximum emission between December 2024 and September 2025. The expected X-ray maximum is less clearly defined than in the previous oscillation, likely as a consequence of the intrinsically irregular flux variability in the X-ray band. In contrast, the UV emission exhibits a more coherent behavior, with a peak occurring at comparable epochs. Taken together, these monitoring campaigns provide tentative confirmation of the predictions for the $\gamma$-ray behavior based on its $\sim$2.1-year periodicity, with consistent behavior observed in the UV band at comparable epochs, but no clear corresponding pattern in the X-rays for the second predicted peak.

\subsection{Periodicity}\label{sec:periodicity_results}
The results of the periodicity analysis are presented in Table \ref{tab:period_results}. In the $\gamma$-ray band, we recover the significant ($>$3$\sigma$) well-established $\sim$2.1-year period, consistent with previous detections widely reported in the literature \citep[e.g.,][]{ackermann_pg1553, penil_mwl_pg1553, penil_24candidates_2025}.

In the UV band (filter "uvv"), we perform two complementary analyses: one using a 30-day binned light curve, matched to the $\gamma$-ray binning to facilitate a direct interband comparison, and another using the unbinned data to preserve the original temporal sampling and evaluate the effect of short-timescale variability. The use of a binned light curve provides advantages for periodicity studies, since binning suppresses part of the short-timescale stochastic variability, reduces the impact of isolated fluctuations and observational scatter, and enhances the visibility of long-term coherent modulations. We use a binning method based on the median value within each bin, which has proven to be effective in previous studies \citep[e.g., ][]{bindu_binning, penil_mwl_i}. In both cases, we recover the same characteristic period ($\sim$2.1 yr), although with markedly different significance: $>$3$\sigma$ for the binned light curve, but only $\sim$1$\sigma$ for the unbinned data. This difference highlights the strong impact of short-lived variability on the significance estimate, despite the stability of the recovered period. These results are consistent with previous studies \citep[][]{penil_mwl_pg1553, magic_pg1553_2024}. 

For the X-ray band, we performed two independent analyses using both the unbinned and the 30-day binned light curves. The unbinned data revealed a non-significant period of $\sim$1.5 years, consistent with the findings of \citet{penil_mwl_pg1553} and \citet{aniello_pg1553_xray_2024}. However, as discussed in $\S$\ref{sec:methodology}, these results may be significantly affected by the intrinsic properties of the X-ray light curve, particularly the presence of gaps and strong flaring states.

With respect to gaps, $\sim$90\% of data is missing from the X-ray light curve when compare with an ideal, evenly daily sampled series in which every time bin would have a measurement. Such a large fraction of missing points severely reduces the ability of all the methods employed to detect genuine periodicity, often leading either to suppressed signals or to spurious detections \citep{sagar_segundo, penil_gaps_2025}. As a result, the apparent period could simply be an artifact of the gap distribution in the light curve. To evaluate this hypothesis, we use the test used in \citet{sagar_pg1553, sagar_segundo, penil_gaps_2025}. This approach involves simulating 100,000 synthetic light curves using the method described in \citet{emma_lc}, maintaining the same PSD, PDF, and gap structure (percentage and temporal structure) as the original light curve. The resulting distribution of recovered periods and significances provides a benchmark to identify potential biases toward certain timescales that might align with the detected signal. In addition, we quantified how often the simulations reproduce the same period–significance combination observed in the real light curve. An occurrence is defined when the recovered period falls within the uncertainty range of the real-data result and its significance exceeds that measured in the observed light curve. As shown in Fig. \ref{fig:gaps_test}, the test for GLSP indicates that the detected period is most likely driven by the gap structure, with a $\sim$10\% coincidence rate in the simulations. 

In contrast, the analysis of the binned X-ray light curve yielded different results, with a non-significant hint ($<$2$\sigma$) of a $\sim$2.5-year period. Several factors could explain the absence of a statistically significant $\sim$2.1-year signal in this case. Here, when compared to an ideal evenly sampled series, the fraction of gaps decreases to $\sim$30\%, which lies below the commonly adopted threshold of 50\% above which gaps are expected to strongly bias periodicity searches \citep{penil_gaps_2025}. To further assess the role of gaps, we applied the same test used for the unbinned light curve. The results, shown in Fig. \ref{fig:gaps_test}, indicate that while the gap distribution could in principle contribute to the appearance of the $\sim$2.5-year feature, this explanation is unlikely: the coincidence rate between the observed result and the simulations is 0.5\%. 

Therefore, the dominant factor likely affecting the inference of a $\sim$2.1-year period could be the intrinsic variability of the X-ray emission itself, particularly the impact of high-flux states and strong flaring activity. During these episodes, irregular flux variations can mask or distort underlying periodic signals, leading to a substantial reduction in the apparent significance of long-term oscillations. In extreme cases, both the inferred period and its significance can be affected by uncertainties exceeding 100\% \citep[][]{penil_flares}.

Overall, the combined influence of gaps and flaring states likely explains the weaker and less consistent detection of long-term periodicity in the X-ray band. Despite the tentative indications of a $\sim$2.1-year modulation in the X-ray band, we cannot claim that the X-ray emission of PG 1553+113 follows the same periodic pattern observed at other wavelengths. Extending this effort over longer timescales will allow additional candidate peaks to be observed, providing the statistical leverage needed to clarify the nature of the X-ray variability and to resolve the remaining uncertainties. 

\subsection{Correlation}
The analysis of the correlation between the $\gamma$-ray and X-ray emissions reveals a correlation between the two bands. In particular, we measure a time lag of $-12\pm21$ days, which is consistent with a zero lag, with a significance of 2.8$\sigma$. Given the 30-day binning adopted for the \textit{Fermi}-LAT light curve, time lags within $\pm$30 days remain compatible with a zero-delay scenario. This result is consistent with previous studies, which reported a near-zero lag at comparable significance levels \citep[$\sim$3.0$\sigma$;][]{penil_mwl_pg1553}, and with the findings of \citet{dhiman_xray_corre_pg1553}.

In addition, we restricted the analysis to the X-ray observations obtained during the specific observational windows defined by our \textit{Swift} proposals. Under these conditions, the resulting cross-correlation is again consistent with a zero time lag, with a significance of 2.4$\sigma$, further supporting a contemporaneous variability between the X-ray and $\gamma$-ray emissions during these intervals.

For the cross-correlation between the UV and $\gamma$-ray bands, we obtain results consistent with a zero time lag, measuring a delay of $-11\pm21$ days with a significance of 3.0$\sigma$, in agreement with \citet{penil_mwl_pg1553}. When the analysis is restricted to the peaks intervals covered by our observations, the cross-correlation remains consistent with a $\sim$0-lag but with a reduced significance of 1.7$\sigma$. This decrease in significance is likely related to the most recent $\gamma$-ray oscillation, which exhibits a complex double-peaked structure. When combined with the temporal gaps in the UV coverage, this complexity can dilute the correlation signal and limit the statistical significance of the result.

\begin{figure}
        \centering
        \includegraphics[scale=0.21]{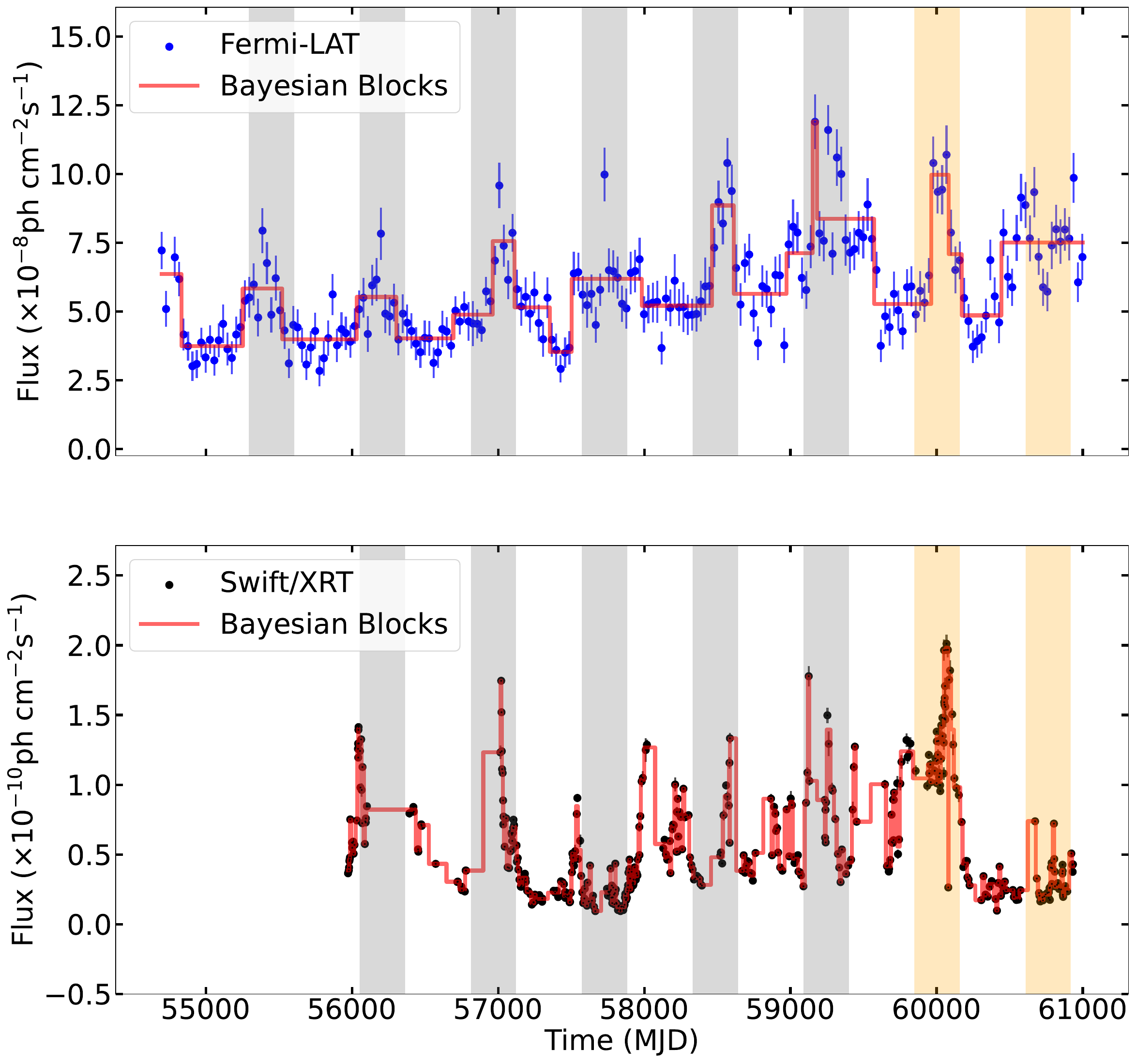}
        \caption{Bayesian blocks of the \textit{Fermi}-LAT and \textit{Swift}-XRT light curves are shown. The gray vertical lines approximately mark the high-emission states corresponding to the inferred period of $\approx$2.1 years, while the orange vertical lines approximately indicate the \textit{Swift} monitoring campaign carried out to track the predicted high states.} \label{fig:bayesian_block}
\end{figure}

We used the Bayesian blocks shown in Fig. \ref{fig:bayesian_block} to highlight the high-emission states of both the $\gamma$-ray and X-ray light curves. In the X-ray panel, most high-emission states appear incomplete, except for the interval coincident with our first observational campaign (Cycle 18; $\S$\ref{sec:obser_camp}). Despite these gaps, most $\gamma$-ray high-emission states have an X-ray counterpart, with two apparent exceptions: the peak around MJD $\sim$58000 and the high-emission state predicted for Cycle 21 ($\S$\ref{sec:obser_camp}). In the UV band, Fig. \ref{fig:bayesian_block_uvot} shows the corresponding counterparts to the $\gamma$-ray high-emission states.

\subsection{Trend Characterization}
\begin{figure}
        \centering
        \includegraphics[scale=0.21]{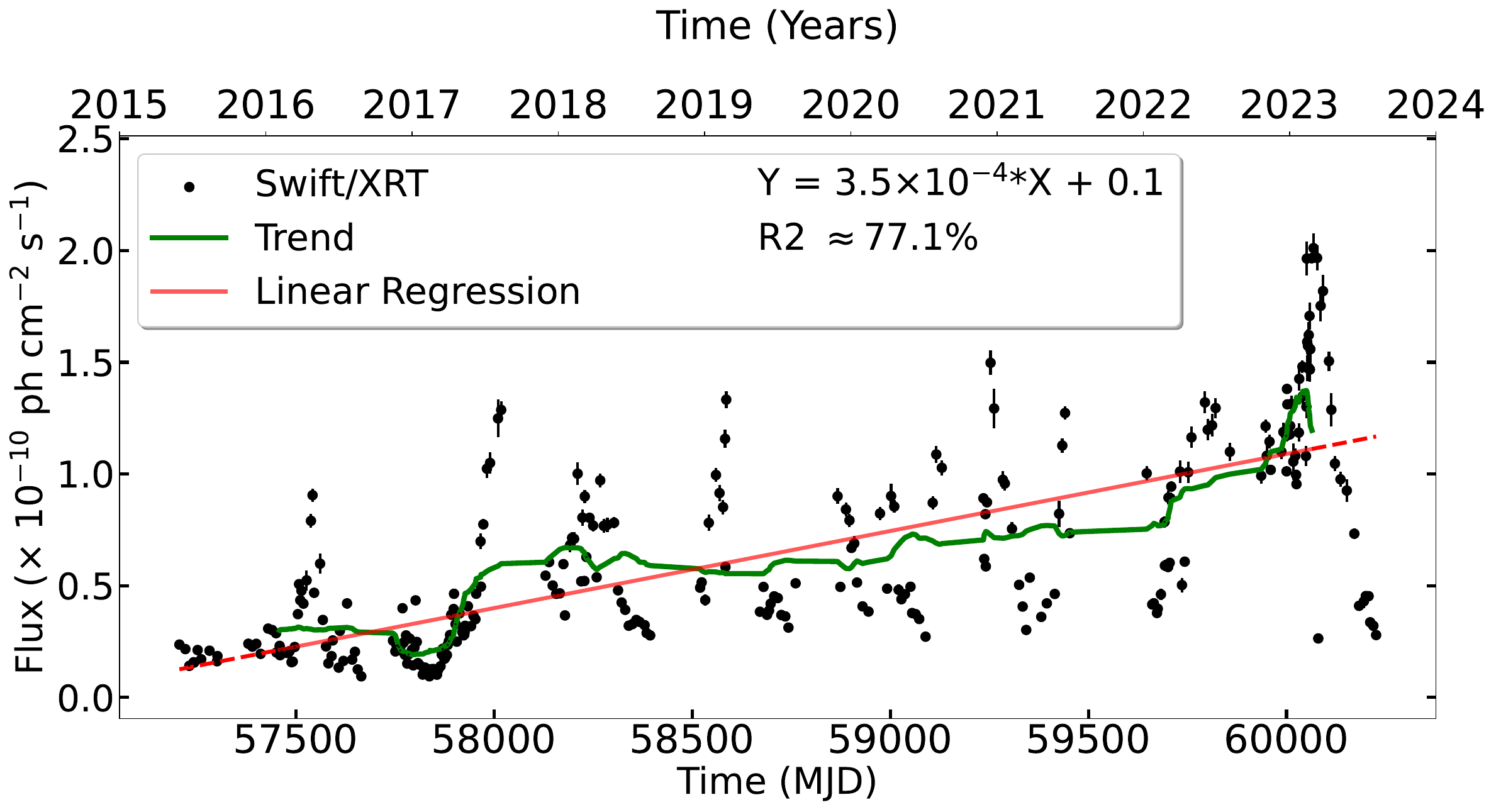}
        \caption{Trend decomposition of the X-ray data. The green line represents the underlying trend extracted by the function \textit{seasonal\_decompose}. Note that the green line covers a shorter time span than the full light curve duration, a result of the trend decomposition applied by this function. The red line indicates the fit of the green line, with the dashed red line extending the fitted line across the entire light curve. R$^{2}$ is 77.2\%, denoting a ''substantial'' fit.} \label{fig:linear_fitting_LC_swift}
\end{figure}

The estimation of the long-term trends in the X-ray band is shown in Fig. \ref{fig:linear_fitting_LC_swift}. The best-fit slope is $\rm{35\pm8\times10^{-5}}$ with R$^{2}$=77.1\%, consistent with the value of $\sim \rm{20\times 10^{-5}}$ reported by \citet{penil_mwl_pg1553}. With the inclusion of the new X-ray observations, this confirms the presence of a persistent upward trend in the X-ray emission of PG 1553+113.

In the $\gamma$-ray band, we obtained a slope of ${70\pm7\times10^{-5}}$, with R$^{2}$=81.3\% (Fig. \ref{fig:trends_fermi_uvot}), which agrees with the results presented by \citet{penil_2025_trend}. For the UV data, the slope of $\rm{20\pm1\times 10^{-5}}$, with R$^{2}$=76.8\% (Fig. \ref{fig:trends_fermi_uvot}), which is likewise compatible with previous results \citep{penil_mwl_pg1553}.

These results demonstrate that all three energy bands, X-ray, $\gamma$ ray, and UV, exhibit linear long-term trends but with different associated slopes. This multi-band agreement strengthens the case for a common underlying physical mechanism driving the gradual brightening of PG 1553+113 as was proposed in \citet{sagar_pg1553}.

\subsection{Relation Power Law Index-Flux}
\begin{figure}
        \centering
        \includegraphics[scale=0.21]{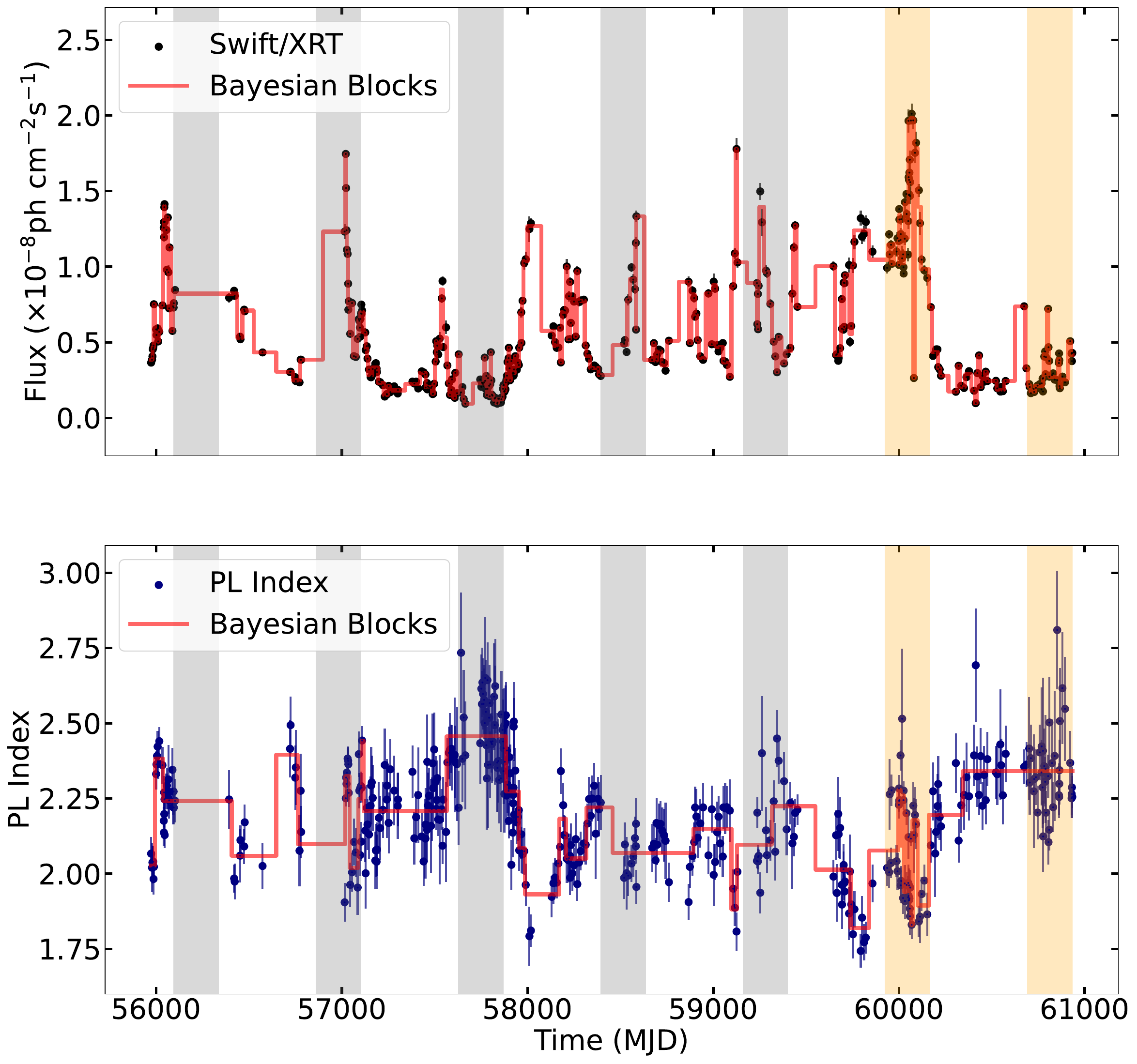}
        \caption{Bayesian block of the X-ray flux and the power-law index resulting from the fitting model. It shows an apparent anti-correlation relation between the flux and the associated PL index. The gray vertical lines mark the high-emission states corresponding to the inferred period of $\approx$2.1 years, while the orange vertical lines indicate the \textit{Swift} monitoring campaign carried out to track the predicted high states.} \label{fig:bayesian_block_plindex_flux}
\end{figure}

Finally, we also performed a study of the X-ray flux and the  photon index (see $\S$ \ref{sec:xray_flux}). Both the X-ray flux and the photon indices are shown in Fig. \ref{fig:plindex_flux}. To investigate possible relationships between these quantities, we applied a Bayesian blocks analysis to both the X-ray flux and the associated photon index. Figure \ref{fig:bayesian_block_plindex_flux} shows an apparent anti-correlation between the X-ray flux and the photon index. This behavior becomes more evident when considering the two monitoring cycles separately: during the high-emission flare of Cycle 18, the PL index exhibits a clear depression (e.g. harder photon index for brighter flux state), whereas in Cycle 21 the X-ray flux displays relatively lower emission while the PL index reaches high (softer) values.

To further quantify this, we performed a \textit{z}-DCF correlation analysis between the two datasets, finding an anti-correlation consistent with a near-zero time lag (-7$\pm$12 days), with a significance of 2.2$\sigma$.

To complement the \textit{z}-DCF analysis, we examined the relationship between the X-ray flux and the photon index using Spearman’s rank correlation coefficient \citep[][]{spearman_2010}, $\rho_{\rm S}$. Spearman’s $\rho_{\rm S}$ can be applied to irregularly sampled time series, provided the two observables are evaluated as one-to-one paired measurements at common epochs. For the X-ray flux and photon-index light curves we obtain $\rho_{\rm S}$=-0.6209. We assessed the significance with Monte Carlo simulations by constructing the null distribution of $\rho_{\rm S}$ from $M$=$10^{5}$ pairs of surrogate light curves that reproduce the sampling pattern and variability properties of the data. The surrogate distribution is centered near zero, with median $\tilde{\rho}{\rm S,\,surr}$=$-2\times10^{-4}$ and a central 68\% interval of $[-0.0668,\ 0.0663]$, implying that chance correlations under the adopted null model typically satisfy $|\rho{\rm S}|\lesssim 0.07$. The observed anti-correlation exceeds all surrogate realizations, yielding a two-sided surrogate-based $p$-value of $p_{\rm surr}$=$10^{-5}$, and demonstrating that the flux--index anti-correlation is highly significant relative to the red-noise hypothesis.

\section{Discussion}\label{sec:discussion}
One of our main results is the confirmation of our predictions for the next high-emission states of PG~1553+113 for the $\gamma$ rays. In our proposals, we forecast two $\gamma$-ray maxima, March 2023 (Cycle 18) and May 2025 (Cycle 21), with an uncertainty of one month. As shown in Fig. \ref{fig:mwl_pg1553}, the $\gamma$-ray light curve displays peaks near May 2023 and June 2025, in agreement with these forecasts. The UV emission shows consistent behavior as well. This outcome highlights the value of using forward predictions as an independent test of candidate periodic patterns. The rationale behind this approach is straightforward: a stochastic process, by definition, cannot be reliably predicted beyond its statistical properties, whereas a genuine periodic signal should allow for reproducible forecasts of future variability. If the predicted maxima or minima of the emission are confirmed by subsequent observations, this provides strong evidence that the variability is not simply the product of noise or chance fluctuations. In particular, it motivates the design of targeted observational campaigns aimed at monitoring epochs when the periodic model forecasts significant changes in emission. 

In the X-ray band, some peaks appear to coincide with the $\gamma$-ray and UV activity, while for others the correspondence is uncertain, particularly for the high-emission state targeted in the Cycle 21 proposal. As a result, establishing the same periodic behavior in X-rays remains challenging. However, we confirmed the presence of the long-term trend in X-ray emission. The detection of consistent behavior across multiple wavelengths suggests a common origin for the MWL emission and, consequently, the action of the same physical mechanism driving the variability. The trend itself lends additional support to the so-called “lump” scenario proposed for PG 1553+113. Nevertheless, further investigation is needed to fully assess the presence of the putative 22-year modulation.

The observed anti-correlation between the X-ray flux and the photon index in PG 1553+113 is consistent with the well-known harder-when-brighter behavior commonly observed in HBL BL Lac objects \citep[e.g.,][]{wang_harder_softer_2018}. Since the X-ray emission in this source is dominated by synchrotron radiation from the highest-energy electrons in the jet, an increase in flux accompanied by a hardening of the photon index can be interpreted as enhanced particle acceleration or fresh injection of high-energy electrons, which temporarily hardens the electron energy distribution and shifts the synchrotron peak to higher energies \citep[][]{haocheng_harder_brighter_2015}. This behavior has been previously reported for PG 1553+113 in multiwavelength campaigns \citep[][]{magic_pg1553_2024}. The near-zero time lag between the flux and spectral variations further supports a scenario in which both quantities are driven by closely coupled acceleration and cooling processes within the same emitting region \citep[][]{haocheng_harder_brighter_2015}. In addition to that, theses highest-energy electrons in the jet have short radiative cooling and acceleration timescales and are therefore highly sensitive to localized and transient dissipation processes, such as magnetic reconnection or turbulent acceleration \cite[][]{petropoulou_radiative_cooling_2018, christie_radiative_cooling_2019}. As a result, the X-ray light curve is often dominated by rapid, large-amplitude fluctuations that introduce strong red-noise variability \citep[][]{vaughan_red_noise}. This enhanced stochastic component can substantially reduce the coherence and statistical significance of any underlying multi-year modulation in the X-ray band. In contrast, the $\gamma$-ray emission may arise from a more spatially extended or temporally averaged particle population within the jet, or from emission zones less affected by rapid cooling \citep[][]{ghisellini_jet_structure_2005, marscher_regions_jet_2014}, allowing dynamical modulations to remain detectable. Under these conditions, correlated high-emission states can coexist with a $\gamma$-ray periodicity and an apparently non-periodic X-ray behavior, either because the periodic component is sub-dominant in X-rays or because the X-ray emission originates from a more intermittent, multi-zone substructure than the $\gamma$-ray emission.

Consequently, given the strong stochastic variability that characterizes the X-ray band, coordinated observational campaigns guided by predictions of future high-emission states are crucial since they may provide a more effective strategy for testing whether the periodic behavior observed in the $\gamma$-ray and UV bands is also present in X-rays. If such a connection is confirmed with more evidences, it would strongly support the presence of a common physical mechanism driving the variability across all these bands. Specifically, the coupling between the $\gamma$-ray and X-ray bands would point to a scenario in which both components arise from the same leptonic population of relativistic electrons. In this framework, the X-ray emission is produced via synchrotron radiation, while the $\gamma$-ray photons originate from the upscattering of these synchrotron seed photons through the synchrotron self-Compton (SSC) mechanism \citep[e.g.,][]{abdo_bllac_corr}. The correlated behavior between $\gamma$ rays and the UV bands further supports this picture, since the UVOT flux traces the high-energy end of the synchrotron component that seeds the SSC process. Therefore, simultaneous or near-simultaneous variability in the UV, X-ray, and $\gamma$-ray domains naturally emerges from the dynamical evolution of a single electron population within a common emitting zone.

\section{Summary} \label{sec:summary}
In this work, we investigated the X-ray emission of PG 1553+113, a well-studied blazar and a leading candidate for hosting a supermassive black hole binary. Our aim was to assess whether the X-ray band reproduces the variability patterns, periodic modulation and long-term trend, previously reported at other wavelengths. To this end, we carried out two dedicated monitoring campaigns targeting predicted high-emission epochs using \textit{Swift}-XRT. We did not find significant evidence for the $\sim$2.1-year period in X-rays. However, targeted observational campaigns during predicted high-emission states, based on the $\gamma$-ray periodicity, revealed hints of a $\sim$0-lag correlation associated with X-ray flaring episodes. In addition, we confirmed the presence of a long-term trend in the X-ray band, consistent with results from other energy ranges. Regarding UV, we found significant results of the $\sim$2.1-year ($\sim$3.0$\sigma$) oscillations and $\sim$0-lag cross correlation with the $\gamma$ rays ($\sim$3.0$\sigma$). These findings strengthen the hypothesis of a common physical origin for the MWL emission of PG 1553+113. Nevertheless, confirming the $\sim$2.1-year periodicity in X-rays will require continued monitoring guided by predictions of future oscillations since such confirmation of periodic pattern by applying natural periodicity-search strategies are not robust due to the intrinsic high-variable and irregular-distributed X-ray emission. 

\begin{acknowledgements}
P.P. and M.A. acknowledge funding under NASA contract 80NSSC20K1562. J.O.-S. acknowledges founding from the Istituto Nazionale di Fisica Nucleare (INFN) Cap. U.1.01.01.01.009. 

This work was supported by the European Research Council, ERC Starting grant \textit{MessMapp}, S.B. Principal Investigator, under contract no. 949555, and by the German Science Foundation DFG, research grant “Relativistic Jets in Active Galaxies” (FOR 5195, grant No. 443220636).
\end{acknowledgements}

\bibliographystyle{aa} 
\bibliography{literature.bib}

\begin{appendix}

\section{Tables}
This section presents the tables that contain the results of different analyses of this study.
Table \ref{tab:period_results} shows the periodicity results. 
\begin{table*}
\centering
\caption{List of periods and uncertainties (top) with their associated local significance (bottom) for the different energy bands analyzed. Following the approach of \citet{penil_mwl_pg1553,penil_24candidates_2025}, and considering a sample size of 1492 objects \citep{penil_2025_4fgl}, the global significance associated with these local significance values is $\sim$0$\sigma$. For X-ray and UV, we include the 30-binned and no binning LC. \label{tab:period_results}}
{%
\begin{tabular}{l|cccccccccc}
\hline
\hline
Energy Band & GLSP & SSA & PDM \\	
\hline
$\gamma$ ray & $2.1^{\pm0.2}_{3.6\sigma}$ & $2.1^{\pm0.1}_{2.4\sigma}$ & $2.1^{\pm0.2}_{4.0\sigma}$ \\
X-ray & $1.5^{\pm0.2}_{0.5\sigma}$ & $1.4^{\pm0.8}_{0.9\sigma}$ & $2.9^{\pm0.8}_{1.4\sigma}$ \\
X-ray (30-day binned) & $2.5^{\pm0.3}_{1.4\sigma}$ & $1.9^{\pm0.6}_{0.9\sigma}$ & $2.5^{\pm0.3}_{1.9\sigma}$ \\
UV & $2.2^{\pm0.2}_{3.0\sigma}$ & $2.3^{\pm0.5}_{0.6\sigma}$ & $1.2^{\pm0.3}_{0.4\sigma}$ \\
UV (30-day binned) & $2.1^{\pm0.2}_{3.7\sigma}$ & $2.1^{\pm0.2}_{2.4\sigma}$ & $2.1^{\pm0.2}_{4.0\sigma}$ \\
\hline
\hline
\end{tabular}%
}
\end{table*}

\section{Figures}
\begin{figure*}
        \centering
        \includegraphics[scale=0.51]{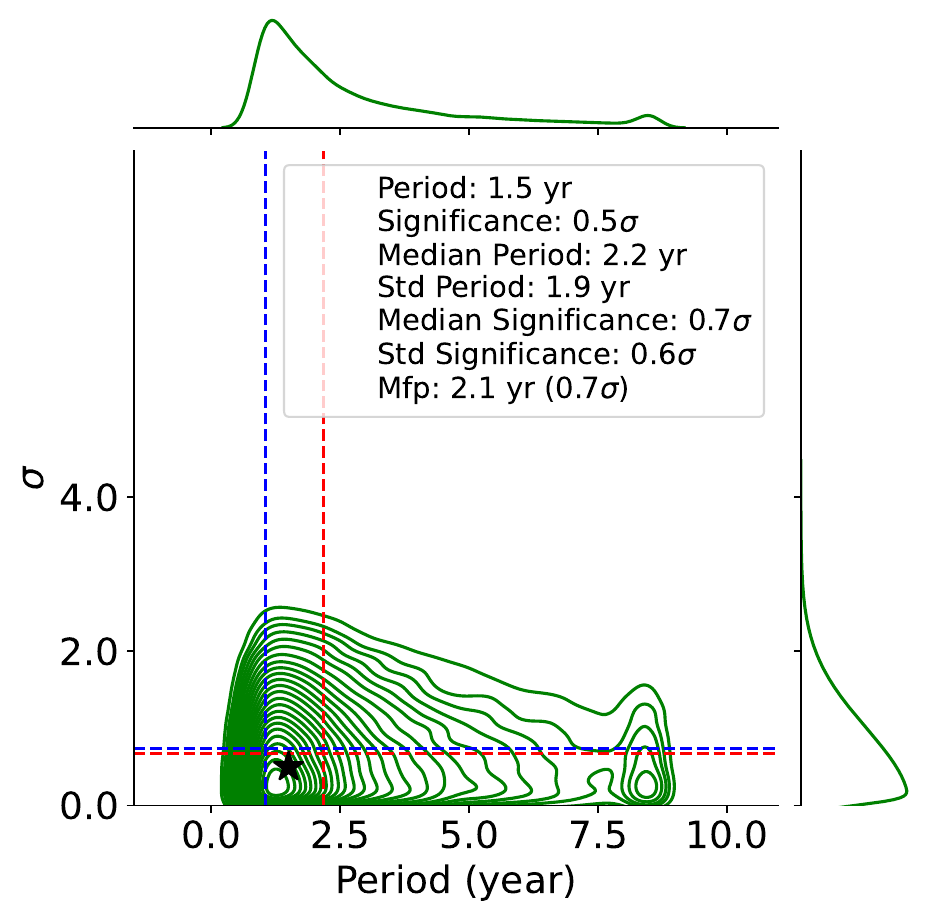}
        \includegraphics[scale=0.51]{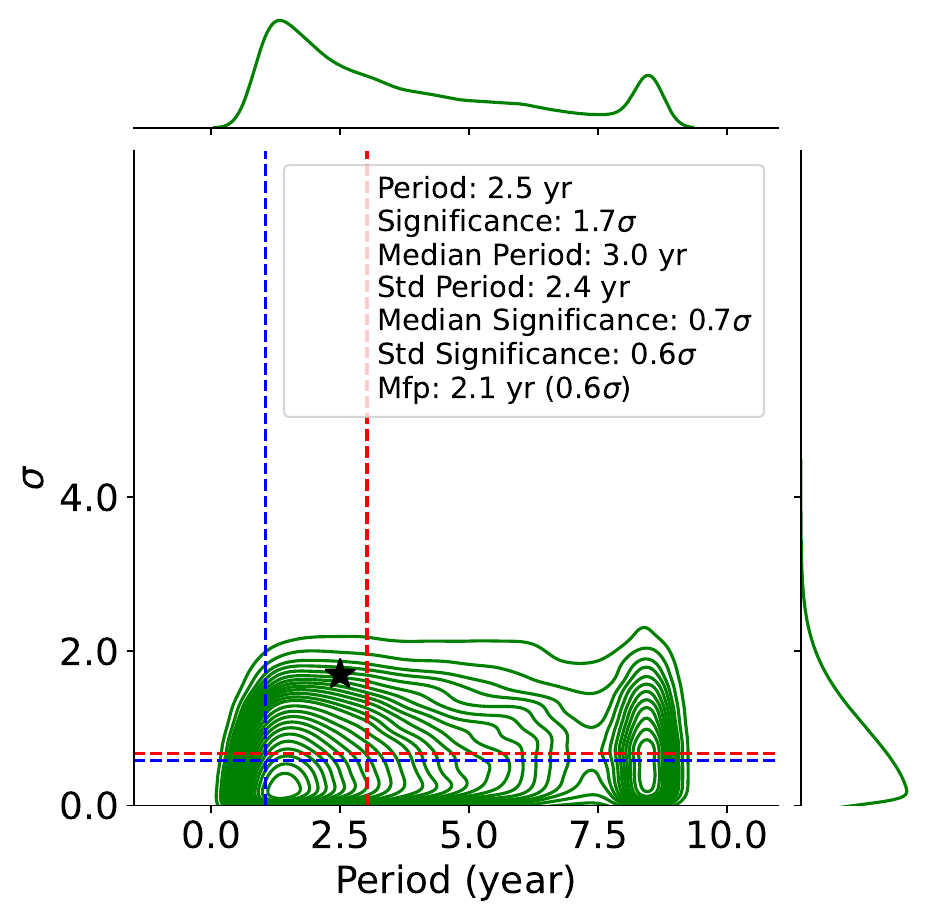}
        \caption{Left: Distributions for the period and significance for the simulated light curves with the same properties as the X-ray data using 100.000 artificial light curves with the same properties as the original light curve (i.e., sampling, gap distribution, PSD, PDF) for the method GLSP. The results denote that the period and significance obtained are compatible with random light curves with the same gap structure as the original X-ray light curve. 
        Right: Distributions for the period and significance for the simulated light curves with the same properties as the 30-day binned X-ray data using 100.000 artificial light curves with the same properties as the original light curve (i.e., sampling, gap distribution, PSD, PDF) for the method GLSP. The results indicate that the period and significance recovered from the real data could, in principle, arise from the gap structure, but this scenario is unlikely in the case of the binned light curve.              
        The dotted red vertical and horizontal lines indicate the median values for both the period and the significance of the test. The blue dotted vertical line highlights the most frequently occurring period in the tests (and the associated significance), emphasizing its prominence in the distribution. The ``Median Period'' represents the median of all periods resulting from the test, and the ``Std Period'' is the standard deviation of such periods' distribution. The ``Median Significance'' represents the median of the significance distribution associated with the test, and the ``Std Significance'' is the standard deviation of this significance distribution. ``Mfp'' represents the most frequent period resulting from the test.} \label{fig:gaps_test}
\end{figure*}

\begin{figure*}
        \centering
        \includegraphics[scale=0.22]{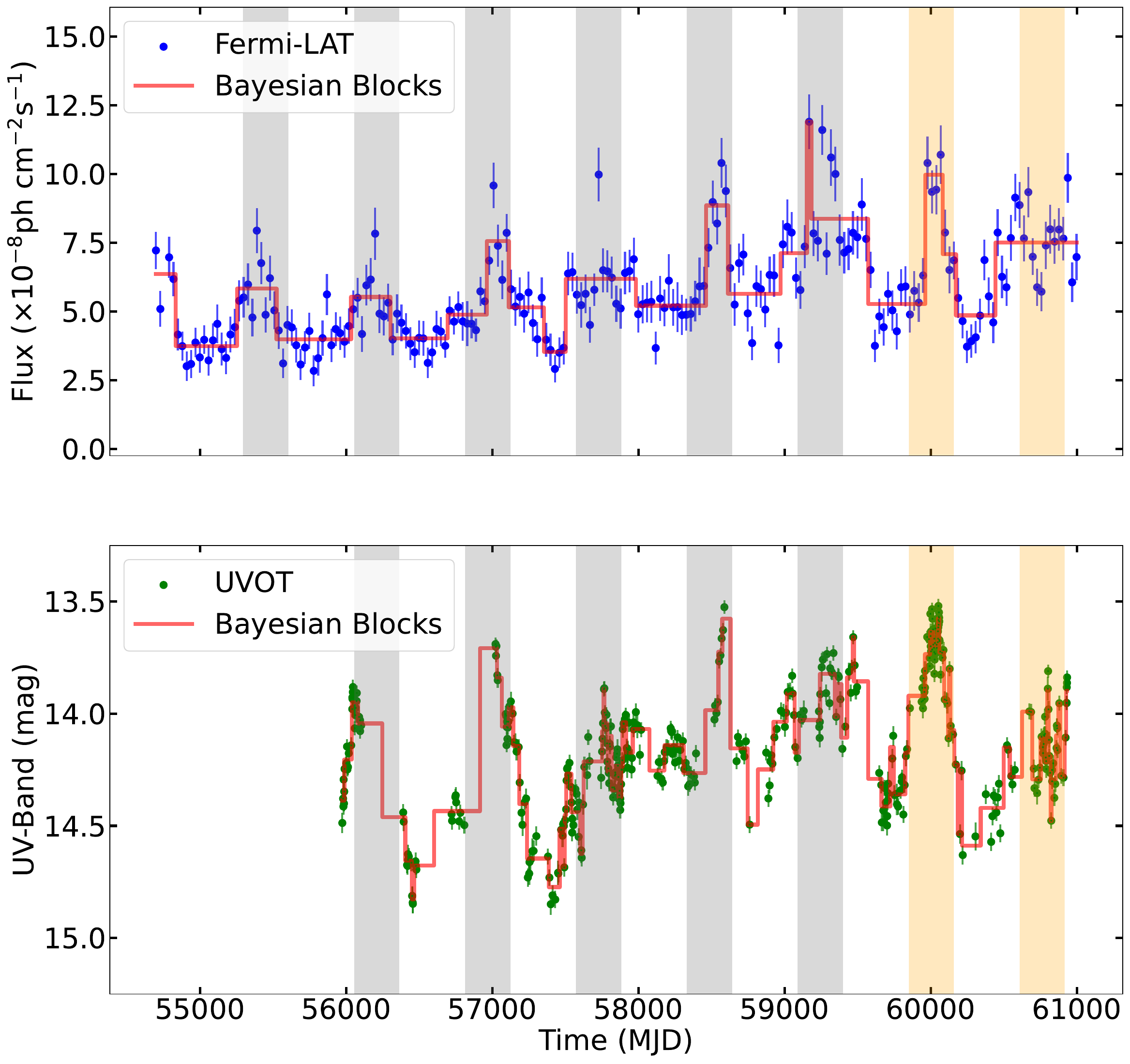}
        \caption{Bayesian blocks of the Fermi-LAT and UVOT (filter "uvv") light curves are shown. The gray vertical lines approximately mark the high-emission states corresponding to the inferred period of $\approx$2.1 years, while the orange vertical lines approximately indicate the \textit{Swift} monitoring campaign carried out to track the predicted high states. The figure highlights the near-coincidence of the $\gamma$-ray high-emission phase with a similar state observed in UV.} \label{fig:bayesian_block_uvot}
\end{figure*}

\begin{figure*}
        \centering
        \includegraphics[scale=0.22]{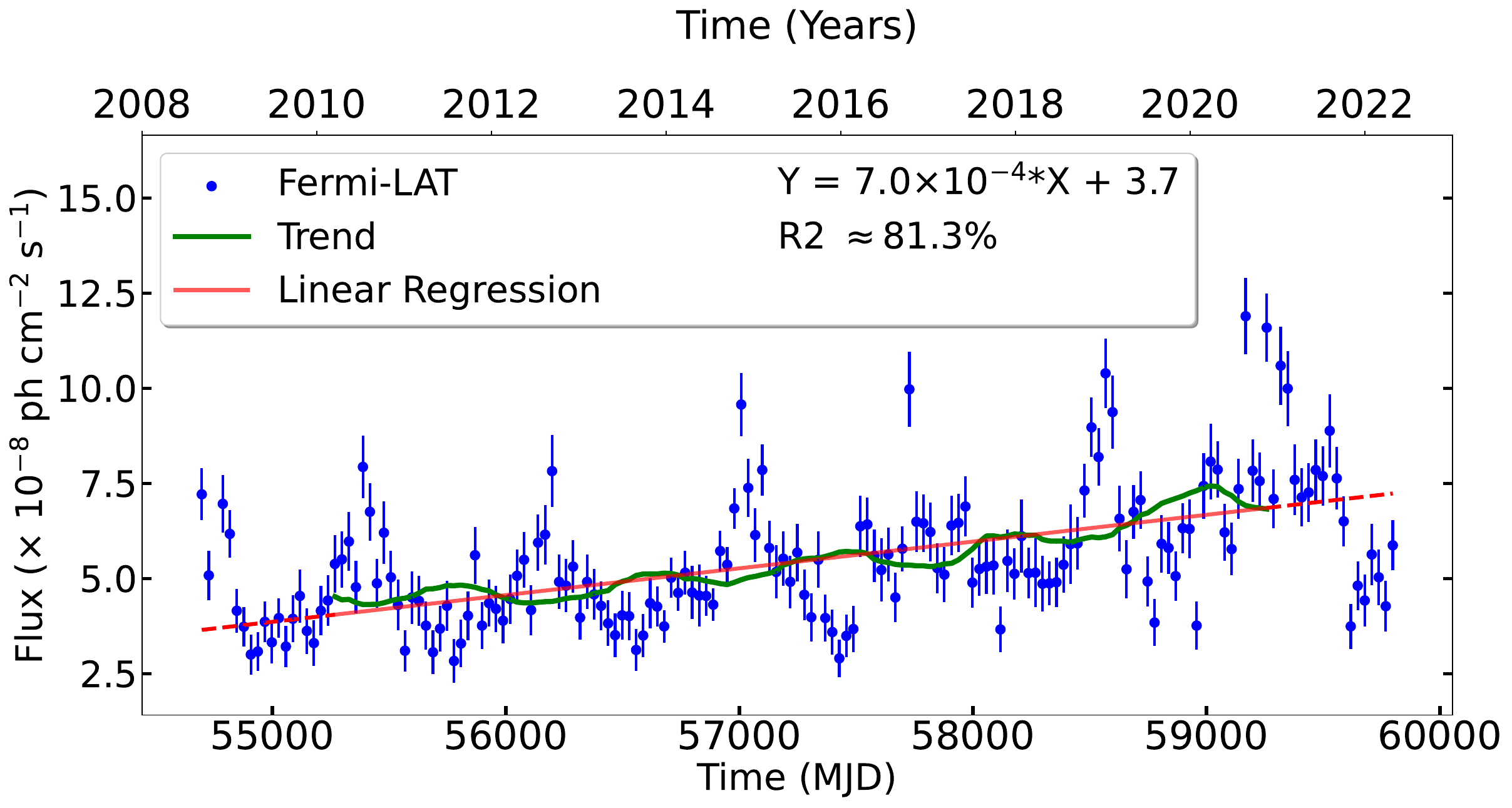}
        \includegraphics[scale=0.22]{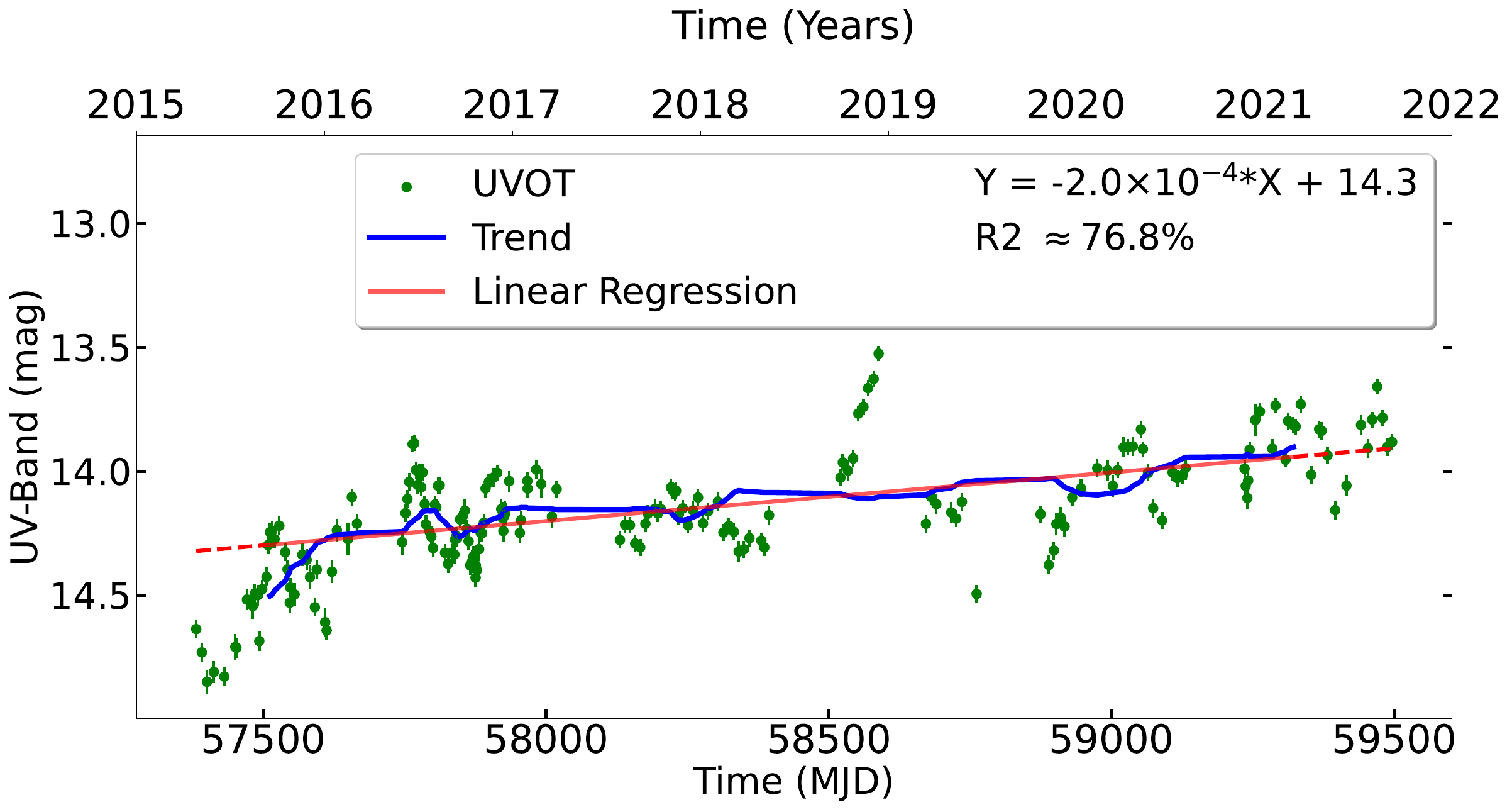}
        \caption{Trend decomposition. Left: $\gamma$ ray. Right: UV (filter "uvv"). The green (blue) lines represent the underlying trend extracted by the function \textit{seasonal\_decompose}. Note that the green (blue) line covers a shorter time span than the full light curve duration, a result of the trend decomposition applied by this function. The red line indicates the fit of the green line, with the dashed red line extending the fitted line across the entire light curve. R$^{2}$ is 81.3\% and 76.8\% for $\gamma$ ray and UV, respectively,  denoting a ''substantial'' fit in both cases.} \label{fig:trends_fermi_uvot}
\end{figure*}

\begin{figure*}
        \centering
        \includegraphics[scale=0.17]{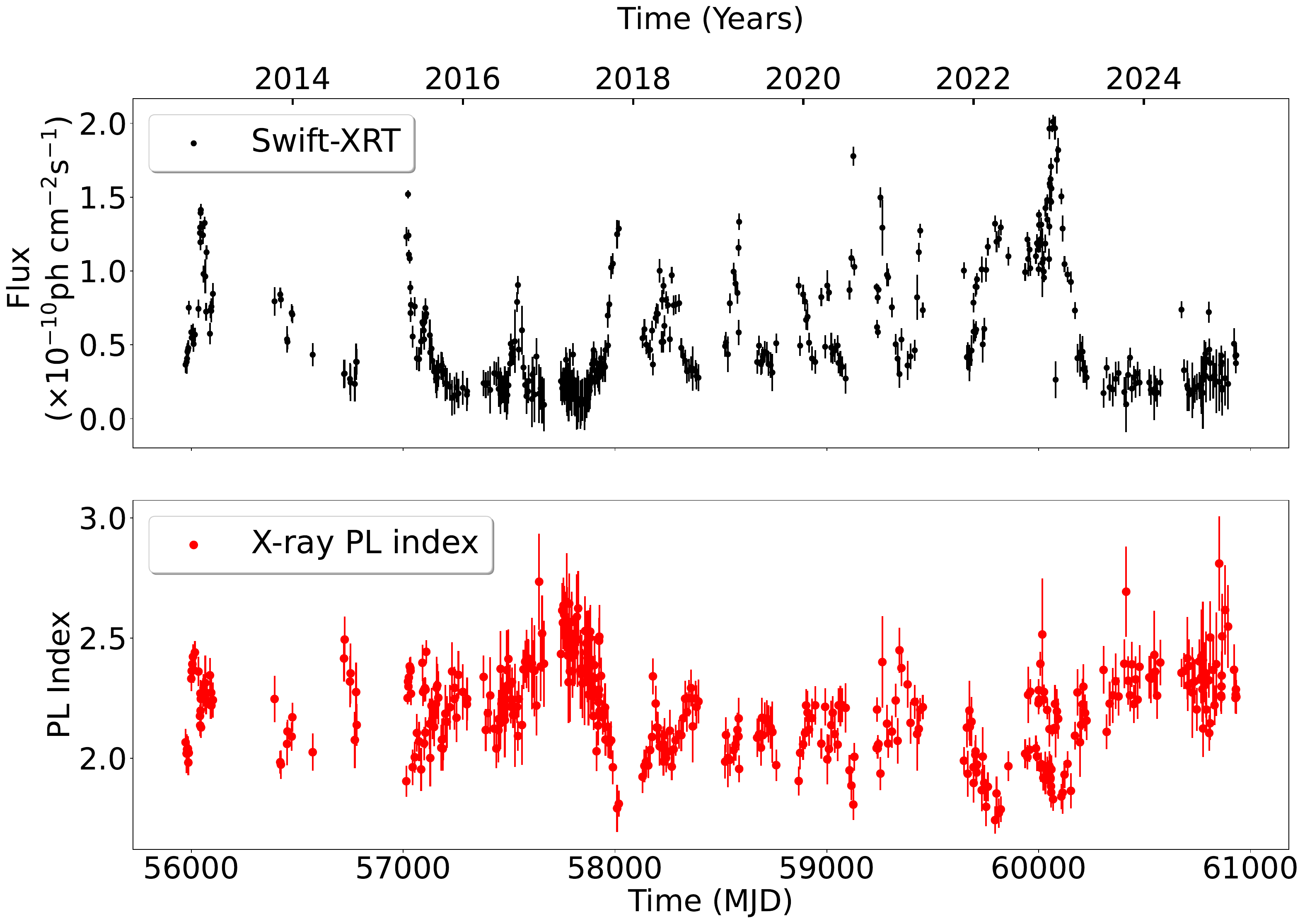}
        \caption{Representation of the X-ray flux and the power-law indices resulting from the fitting model.} \label{fig:plindex_flux}
\end{figure*}

This section presents figures corresponding to different parts of the analysis. Figure \ref{fig:gaps_test} shows a test to evaluate if the period-significance obtained from the X-rays could result from the gaps. Figure \ref{fig:bayesian_block_uvot} shows the Bayesian block representations of the \textit{Fermi}-LAT and UVOT data (filter "uvv"). Figure \ref{fig:trends_fermi_uvot} shows the trend  study for the $\gamma$-rays and the UV bands. Figure \ref{fig:plindex_flux} shows the X-ray curve and the associated power-law indices. 

\section{Discarded \textit{Swift} Observations}\label{appendix:discards}

This section presents details and figures relating to discarded \textit{Swift} observations in the analysis. 

\begin{figure*}
        \centering
        \includegraphics[scale=0.21]{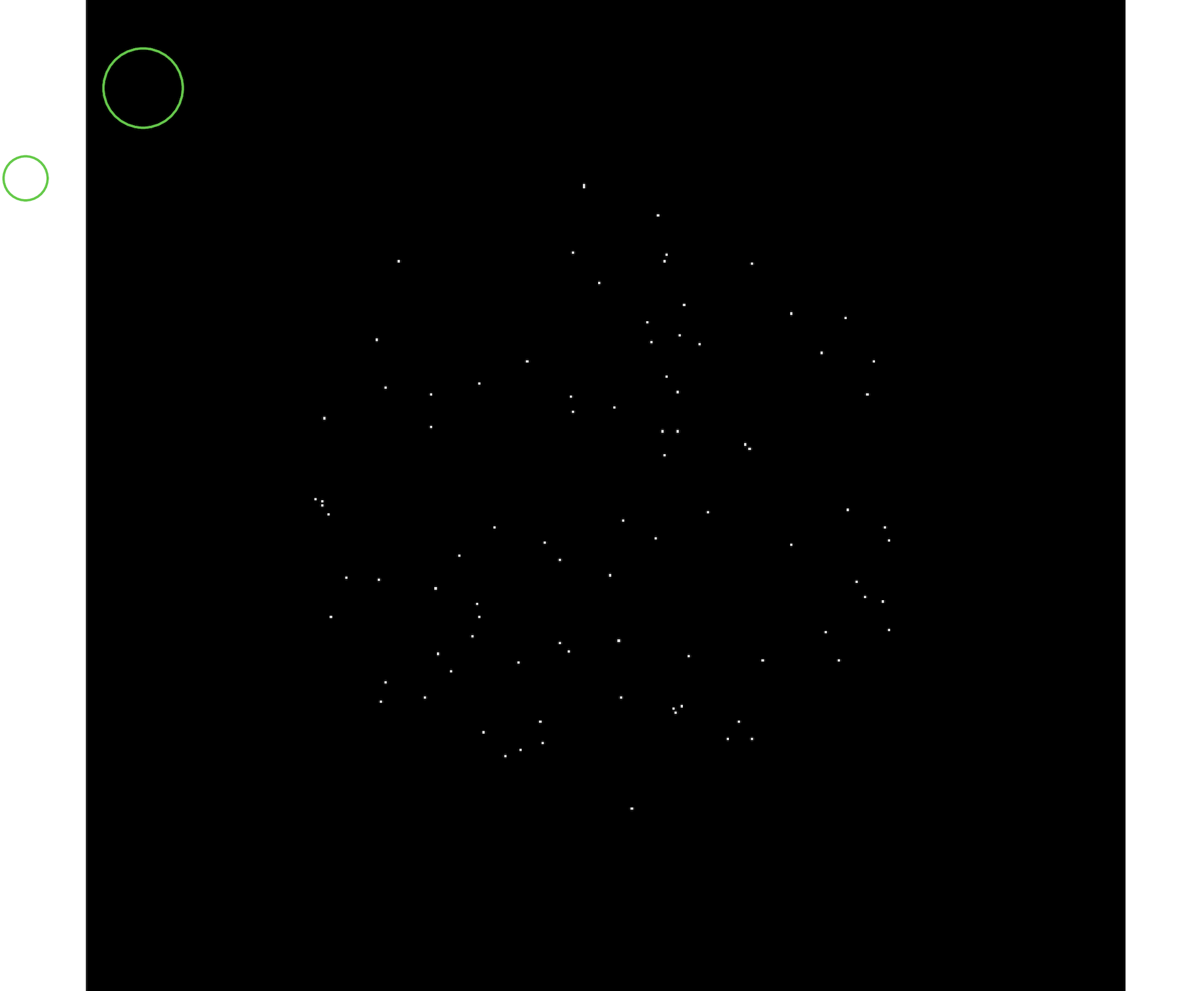}
        \includegraphics[scale=0.3]{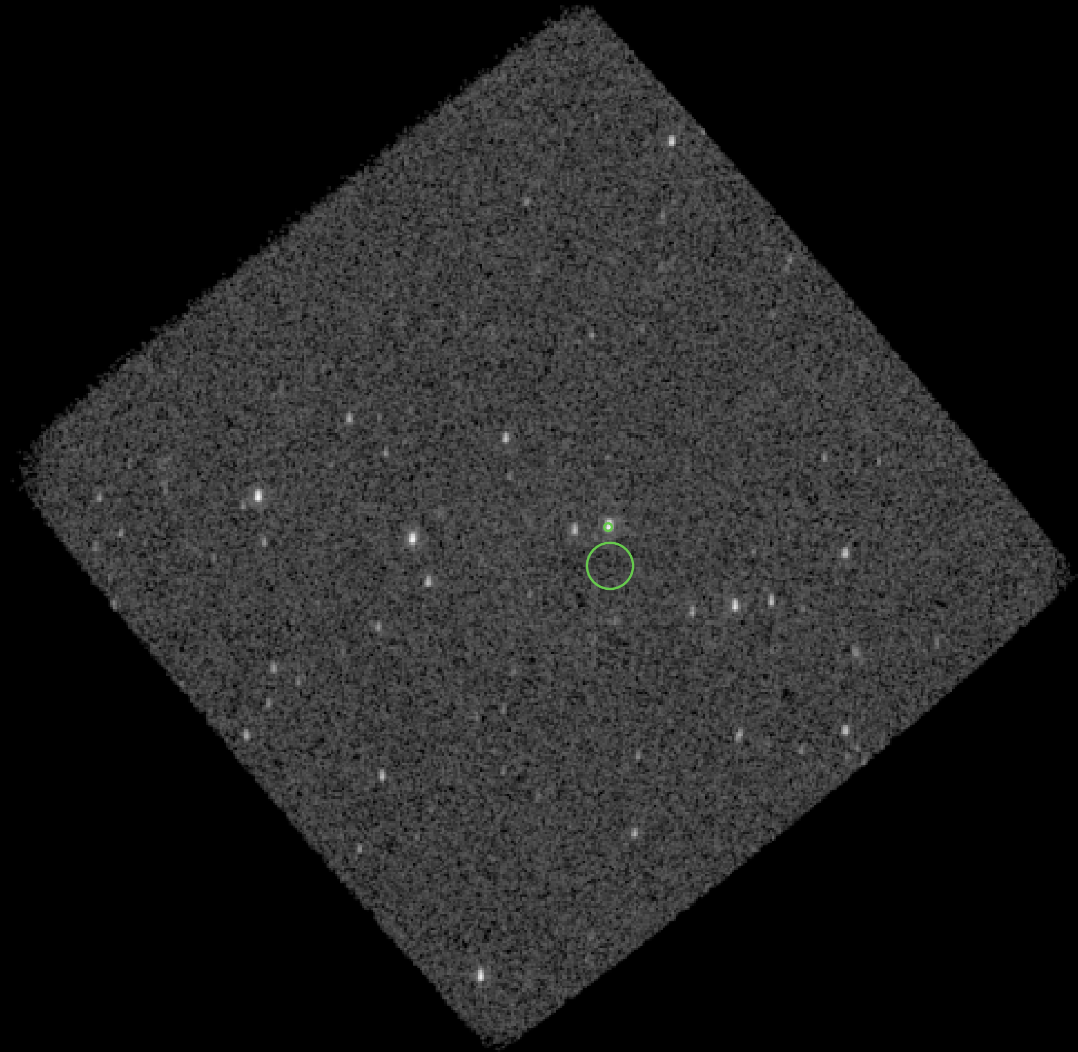}
        \includegraphics[scale=0.3]{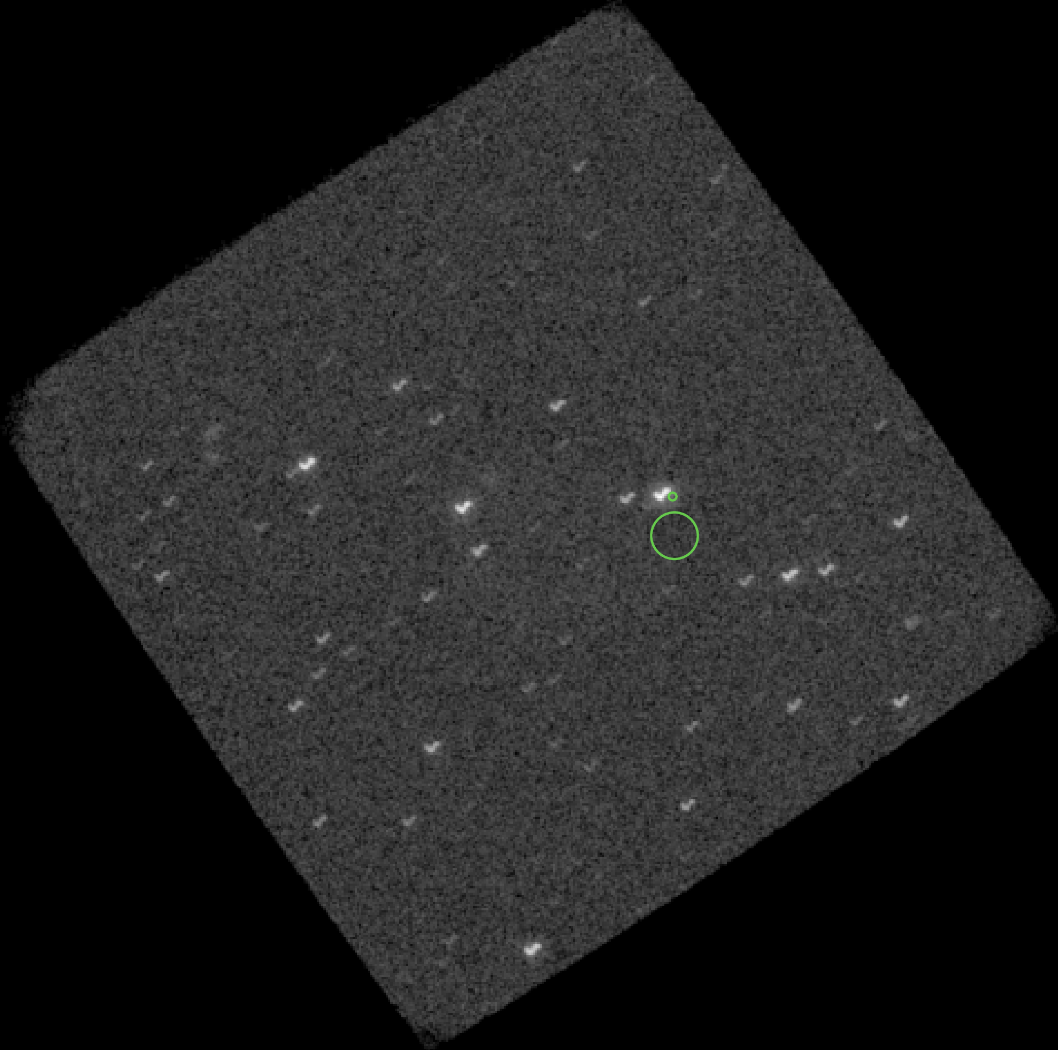}
        \caption{Examples of discarded \textit{Swift} observations after a visual inspection of the FOV and the extraction (small) and background (large) regions. \textit{Left:} Discarded XRT pc observation (OBSID 00046686005), due to incorrect coordinate association. Incidentally, the source is not detected in this observation, implying that the data would have been discarded at a later stage, when the fit would not have converged. \textit{Center and Right:} Examples of discarded UVOT observations (OBSIDs 00031368135 and 00031368090, respectively), due to varying degrees of imperfect tracking.} \label{fig:xrt_obs_problems}
\end{figure*}

The \textit{Swift}-XRT analysis results only on the discarding of three observations, one of them due to bad coordinate association (OBSID 00046686005, see Fig. \ref{fig:xrt_obs_problems}), and two of them due to empty event files (OBSIDs 00019035008 and 00046686005).

The \textit{Swift}-UVOT analysis results in the removal of 28 observations, all of them due to varying degrees of bad tracking. Two examples, of different levels of tracking loss, are shown in Fig. \ref{fig:xrt_obs_problems}. The list of removed OBSIDs is as follows: 00031368072, 00031368090, 00031368129, 00031368135, 00031368137, 00031368177, 00031368181, 00031368247, 00035021014, 00035021016, 00035021021, 00035021073, 00035021074, 00035021085 00035021133, 00035021144, 00035021155, 00035021176, 00035021212, 00035021215, 00035021221, 00035021223, 00035021224, 00035021225, 00035021226, 00046686005, 00092186008, 00096880004.

\end{appendix}

\end{document}